# Research Leadership Flow Determinants and the Role of Proximity in Research Collaborations

Chaocheng He[a,b]   Jiang Wu[a]   Qingpeng Zhang[b,1]

a.   School of Information Management, Wuhan University, Wuhan, Hubei, China.
b.   School of Data Science, City University of Hong Kong, Kowloon, Hong Kong, China.

**Abstract**

Characterizing the leadership in research is important to revealing the interaction pattern and organizational structure through research collaboration. This research defines the leadership role based on the corresponding author's affiliation, and presents, the first quantitative research on the factors and evolution of five proximity dimensions (geographical, cognitive, institutional, social and economic) of research leadership. The data to capture research leadership consists of a set of multi-institution articles in the fields of "Life sciences & biomedicine", "Technology", "Physical Sciences", "Social Sciences" and "Humanities & Arts" during 2013-2017 from Web of Science Core Citation Database. A Tobit regression-based gravity model indicates that research leadership mass of both the leading and participating institutions and the geographical, cognitive, institutional, social and economic proximities are important factors of the flow of research leadership among Chinese institutions. In general, the effect of these proximities for research leadership flow has been declining recently. The outcome of this research sheds light on the leadership evolution and flow among Chinese institutions, and thus can provide evidence and support for grant allocation policy to facilitate scientific research and collaborations.

**Keyword** Research collaborations, Social network analysis, Collaboration networks, Research leadership, Gravity model, Proximity dimensions

---
[1] Corresponding author: Qingpeng Zhang (qingpeng.zhang@cityu.edu.hk); all authors have equal contributions.

# 1 Introduction

Research collaboration is highly valued by academic communities for that it combines and diffuses complementary knowledge expertise of different background (Padial, Nabout et al. 2010). Research collaborations have been increasingly prevalent in academia (Sun, Wei et al. 2017). Research collaboration is defined as the working together of researchers to achieve a common goal of producing new scientific knowledge (Katz and Martin 1997). It has been recognized that the research collaborations among researchers/institutions/countries form a task-oriented complex systems with various interaction and organizational patterns (Newman 2004).

Characterizing the collaboration patterns have been a focus of scientometrics (Nguyen, Ho-Le et al. 2017). In particular, identifying key proximity factors that promote research collaboration can inform actionable policy making to facilitate effective collaborations that carry out good research, which could eventually enhance the prospects of science as a whole (Scherngell and Hu 2011). Existing research mainly focuses on examining the geographic, socioeconomic, and cognitive proximity as potential determinants of research collaboration using a gravity model (Hoekman, Frenken et al. 2010, Plonikova and Rake 2014, Sidone, Haddad et al. 2017). To summarize, the intensity of research collaboration is greater if researchers are closer to each other geographically (physical distance between the two researchers/institutions/countries) (Fernandez, Ferrandiz et al. 2016, Jiang, Zhu et al. 2018); socioeconomic factors such as educational R&D are positively correlated with research activities (Wagner, Park et al. 2015); the cognitive factors such as the similarity in research topics also influence the intensity of research collaboration (Gilsing, Nooteboom et al. 2008). Therefore, understanding the proximity effect is important to characterize the research collaborations.

The previous studies on research collaboration and its proximity factors have three limitations. First, the collaboration relationship was homogeneously assigned among authors for the same article, without considering the research leadership. However, the collaboration relationship between two non-corresponding authors is different from the relationship between a co-author and the corresponding author, who is the leader of the research (Jarneving 2010, Wang, Xu et al. 2013) . Typically, the first author is usually an early-career scientist who undertakes the research (e.g. a PhD student or postdoc); the corresponding author is usually the senior person who supervises and shapes the study (e.g. supervisor) (Sekara, Deville et al. 2018). Other co-authors are usually the researchers who assist the first and corresponding authors in the carrying out the work or writing the paper (not necessarily in the same group) (Hu, Rousseau et al. 2010). Therefore, the common practice is to define the first and corresponding author as the research leaders. The relationship between co-authors and research leaders are stronger than those among non-leaders (Wang, Xu et al. 2013). With the recent trend of the increasing size of multi-institution authorship, the effect of research leadership is becoming even more significant (Wang and Wang 2017). Second, although Boschma (2005) identified five notions of proximity (geographical, cognitive, institutional, organizational, and social), most studies failed to systematically and comprehensively examine the relationships between these key identified factors and research collaboration (Fernández, Ferrándiz et al. 2016, Jiang, Zhu et al. 2018). There are, to the best of our knowledge, only two studies involving all five proximity dimensions, but these two studies used static data without revealing the evolution of these five proximities (Plotnikova and Rake 2014, Fernandez, Ferrandiz et al. 2016). Third, existing studies were mainly based on regional/country-level data, which is with a relatively low resolution of depicting the research collaborations and leadership (Hoekman, Frenken et al. 2010, Gui, Liu et al. 2018). Studying the collaborations at the institutional-level can provide high-resolution of collaboration patterns and inform actionable policy making, and thus is critically needed for characterizing the flow of research leadership.

To fill these research gaps, we extend the literature in several ways. First, we introduce a new measure, research leadership (RL), where we only account for the collaborative relationship between a co-author and the first author or corresponding author. Second, we focus on proximity as one of the main determinants of research collaboration (Gui, Liu et al. 2018, Plonikova and Rake 2014). Following Boschma (2005), we examine the effect of all the four proximity dimensions (geographical, cognitive, institutional, and social proximity) on research collaboration using a gravity model. Note that we do not have organizational proximity since we are focusing on the RL among institutions in specific domains. In addition, we also examine the effect of economic proximity, which has been found to be associated with research performance and collaborations (Hwang 2008, Acosta, Coronado et al. 2011). Third, we examine the research collaboration between Chinese institutions. Because the corresponding authorship is explicitly set to be the primary criterion in the research evaluation system in China (for example, promotion practice and grant review) (Hu, Rousseau et al. 2010).

The paper is organized as follows. Section 2 reviews the literature. Section 3 describes the methods and data. Section 4 and Section 5 present the results and robustness check, respectively. The paper is concluded in Section 6 with discussions of the limitations and future work.

## 2 Literature review

### 2.1 Research collaboration

Research collaboration is a form of intense interaction that allows for effective communication and sharing of capabilities and other resources (Heffner 1981). Research collaboration may lead to various outcomes, such as co-authored paper, patents, deepened personal contact, or nothing at all. Among these outcomes, journal papers have been widely adopted as a reliable proxy to measure the intensity of research collaboration (Newman 2004, Acosta, Coronado et al. 2011). Therefore, the volume of co-authored journal papers is adopted to reflect the degree of intensity of research collaboration.

### 2.2 Research leadership in research collaboration

It has been recognized that the first author and corresponding author often lead the research collaboration in most fields such as biological, engineering, and management (Sekara, Deville et al. 2018). Typically, the person who carries out the research and writes the paper is the first author (Hu, Rousseau et al. 2010). The corresponding author, on the other hand, is responsible for leading the research team, securing grant to sustain the study, shaping the research ideas, designing the roadmap of a project and corresponding with editors, coordinating the collaboration among other co-authors (Wang and Wang 2017). The other co-authors are usually with complementary contributions to assist in undertaking the research. Recently, there is a trend of the increasing size of authorship and more frequent cross-institutional collaborations (Osorio 2018), and thus makes the leadership role of the corresponding author more pronounced (Wang, Wu et al. 2014).

At institution-level, the corresponding author's affiliation is commonly recognized as the leading institution in the project (Alvarez-Betancourt and Garcia-Silvente 2014). Therefore, the corresponding authorship is widely adopted to be a key measure of research contribution and the leadership role (Hu, Rousseau et al. 2010). In many countries/regions (such as China), the corresponding authorship is set to be the primary criterion for research evaluations (for example, promotion and grant review) (Hu, Rousseau et al. 2010). Although the first authorship is also a key indicator of leadership, the first and corresponding authors usually belong to the same institution (for example, the first author is the student/postdoc of the corresponding author) (Wang, Wu et al. 2014). Therefore, we set the corresponding author's affiliation as the leader in cross-institution collaborations.

**2.3 Proximity of research collaboration**

The dynamic interplay of various proximity factors influences the outcome of research collaboration (Hoekman, Frenken et al. 2010). Proximity is an underpinning theoretical framework in investigating the determinants of research collaboration (Berge 2017, Hoekman, Frenken et al. 2010, Gui, Liu et al. 2018, Plonikova and Rake 2014). Boschma (2005) identified five commonly adopted proximities dimensions: geographical, cognitive, institutional, organizational, and social. Recent research has highlighted the significant role that the economic proximity plays during research collaboration. Since we focus on the research leadership between institutions in specific domains, the organizational proximity is not applicable here. The detailed descriptions of selected proximities are as follows.

Geographical distance refers to the physical distance between researchers, which can be measured by absolute metrics such as kilometers, or relative metrics such as travel times (Plotnikova and Rake 2014). There is plenty of evidence showing that the geographical distance between researchers is negatively associated with the intensity of collaborations between them, indicating that researchers/institutions that are close to each other are more likely to collaborate (Hoekman, Frenken et al. 2010).

Cognitive proximity refers to the similarity between researchers' research interests and background. The cognitive proximity partly determines the researchers' absorptive capacity in collaborative projects. The cognitive proximity has been recognized to be positively associated with the intensity of collaborations (Scherngell and Barber 2011). There is also evidence showing that close cognitive proximity does not always benefit research collaboration because of the lack of complementary backgrounds in the research team (Gilsing, Nooteboom et al. 2008, Balland 2012).

Institutional proximity is defined as the similarity between institutions, such as cultural norms and economic development. High institutional proximity could reduce the uncertainty and costs of interactions, and facilitate the establishment of mutual trust (Boschma 2005, Boschma and Frenken 2009). It is usually modeled as a dummy variable to capture whether the collaborators belong to the same administrative division (such as province and state) or the same country. It has been recognized that research collaboration is more frequent among institutions within the same administrative division/ country/ linguistic area (Hoekman, Frenken et al. 2010).

Social proximity refers to the social embeddedness based on the more intangible closeness, such as friendship or prior collaborations between researchers, which is expected to stimulate knowledge interactions and research collaborations owing to the established trust and commitment (Coenen, Moodysson et al. 2004, Boschma 2005). Social proximity is usually measured by the existing/prior collaborations and has been found to be positively associated with future collaborations between individuals, institutions, and countries (Niedergassel and Leker 2011, Hoekman, Scherngell et al. 2013,

Plotnikova and Rake 2014) .

Economic proximity, also named as the socioeconomic proximity, refers to the degree of closeness of academic-related economic resource between institutions (Fernandez, Ferrandiz et al. 2016). According to the center-periphery hypothesis, the difference in academic economic resources between institutions could influence the research collaborations (Schubert and Sooryamoorthy 2009). Economic proximity has been found to have a mixed effect on the collaboration intensity at country level (Hwang 2008, Parreira, Machado et al. 2017, Ling, Zhu et al. 2018). There is few research on examining the influence of economic proximity on the collaboration between institutions.

**Table 1. Summary of Literature in Collaboration Proximity**

| Proximity | Measurement | Remarks and Citation |
|---|---|---|
| Geographical distance | The distance between the capitals of the collaborating countries | It measures the research collaborations at the country level (Gui, Liu et al. 2018, Jiang, Zhu et al. 2018) |
| | The distance between collaborating institutions | It measures the research collaboration at the institution level (Fernandez, Ferrandiz et al. 2016) |
| | The distances between the biggest cities of the two countries | It measures the research collaborations at the country level (Plotnikova and Rake 2014) |
| | The distance between the geographic centers of the regions | It measures the research collaboration at the regional level (Scherngell and Hu 2011, Berge 2017). |
| | The distance between the latitudinal and longitudinal centroid of each country | It measures the research collaboration at the country level (Parreira, Machado et al. 2017). |
| Cognitive proximity | The cosine similarity of the publication counts in 12 disciplines between two institutions | It measures the research collaboration in multiple research fields at the institution level (Fernandez, Ferrandiz et al. 2016). |
| | The cosine similarity of the publication counts in 6 disciplines between two countries | It measures the research collaboration in multiple research fields at the country level (Hoekman, Frenken et al. 2010). |
| | The cosine similarity of the publication counts in 15 sub-areas of therapeutic between two countries | It measures the research collaboration in pharmaceutical research at the country level (Plotnikova and Rake 2014). |
| | The cosine similarity of the publication counts in 22 disciplines between two countries | It measures the research collaboration in multiple research fields at the country level (Gui, Liu et al. 2018). |
| | The simultaneous use of two dimensions: | It measures the university–industry collaboration (Garcia, Araujo et al. 2018). |
| | the cosine similarity of the publication counts in 6 scientific fields and the cosine similarity of the publication counts in 12 industry sectors pairs | |
| Institutional proximity | Dummy variable: 1 if two institutions are in the same country; 0 otherwise. | It measures cross-country research collaboration at institutional level (Hoekman, Frenken et al. 2009, Fernandez, Ferrandiz et al. 2016). |
| | Dummy variable: 1 if two municipalities both have public university campuses; 0 otherwise. | It measures research collaboration between two municipalities in Brazil (Sidone, Haddad et al. 2017). |
| | Dummy variable: 1 if two countries share a common official language; 0 otherwise. | It measures research collaboration at country level (Gui, Liu et al. 2018). |
| | The cosine similarity of 9 indicators in *Institutional Profiles Database* between two countries | It measures research at country level (Plotnikova and Rake 2014). |
| Social proximity | Dummy variable: 1 if two institutions have prior collaborated; 0 otherwise. | It measures research collaboration at institutional level (Petruzzelli 2011, Fernandez, Ferrandiz et al. 2016). |
| | The count of joint-publications by the two countries | It measures research collaboration at country level (Plotnikova and Rake 2014). |
| | The degree of overlap of two countries' prior collaborators | It measures research collaboration at country level (Gui, Liu et al. 2018). |
| | Dummy variable: 1 if two collaborators have a partner in common; 0 otherwise. | The stochastic actor-based model SIENA to model the network dynamic and to estimate parameters (Balland 2012). |
| Economic proximity | Absolute difference in the average Higher education R&D expenditures as % of the GDP. | It measures cross-country research collaboration at institutional level (Fernandez, Ferrandiz et al. 2016). |
| | Absolute difference in per capita higher education R&D expenditure | It measures research collaboration among European universities (Acosta, Coronado et al. 2011). |
| | Absolute difference in real GDP per capita | It measures research collaboration at country level (Jiang, Zhu et al. 2018) |
| | Absolute difference of gross regional product | It measures research collaboration at regional level in China (Scherngell and Hu 2011). |

## 2.4 Impact Measures using Gravity model

Spatial interaction patterns, such as inter-regional/international trade, population commuting, and migration, can be modeled by the analogy to Newton's law of universal gravitation (Burger, Van Oort et al. 2009). The basic idea of the gravity model stems from the law of universal gravitation, hypothesizing that the gravitational force between two objects is directly related to the mass of the objects, and is inversely proportional to the physical distance between them. Initially, the gravity model has been successfully used in predicting trade flows between countries (Isard 1954, Anderson 1979). The model has been applied to a variety of fields. In particular, because the gravity model can capture the joint-effect of both the mass (for example, the number of publications) and distance/proximity (for example, physical distance) of entities, it has been extensively applied to modeling research collaborations (Acosta, Coronado et al. 2011, Parreira, Machado et al. 2017). In this paper, we will integrate the five measures of proximity to measure their impact on RL flows using a gravity model.

## 3 Methods and Data

### 3.1 Measurement of research leadership

The intensity of research collaboration captures the frequency of the research collaboration between two corresponding institutions. In the literature, there are mainly two measures of the intensity of research collaborations: the full count and the fractional count (Berge 2017). The full count is 1 if a paper at least one coauthor from institution $a$ and at least one coauthor from institution $b$, while the fractional count is $2/(N-1)N$, where $N$ is the number of institutions. Following Jiang, Zhu et al. (2018), we used the fractional count to measure research collaboration intensity as it takes the contribution into consideration, instead of simply measuring the participating frequency. Because we defined the role of leadership, there exists the relationship between leaders and others being led to conduct the research. Therefore, the RL is directed (from the leading institution to non-leading participating institutions) and weighted (on the frequency of prior collaborations). We define that a paper possesses a total leadership mass of 1, and the RL flows from the leading institution (the affiliation of the corresponding author) to all other participating institutions (the affiliations of co-authors). Since we focus on the flow of RL, we do not consider a complete network of coauthors; instead, we consider a star-like network (Figure 1) with the leading institution in the center and participating institutions connecting to it (including a self-loop from the leading institution to itself). If there are multiple leading institutions (a paper with multiple corresponding authors affiliated with multiple institutions), the RL mass contributed by them is evenly distributed (Figure 1). In our data, papers with multiple leading institutions account for 3%. The RL flow intensity $C_{ab,i}$ from the leading institution $a$ to institution $b$ in the paper $i$ is expressed as

$$C_{ab,i} = \frac{1}{LIN_i} \times \frac{1}{N_i}, \qquad (1)$$

where $LIN_i$ is the number of leading institutions in paper $i$. And $N_i$ is the number of institutions in paper $i$. Thus, the RL mass contributed by the leading institution is the total RL flow minus the self-loop:

$$LM_{a,i} = \sum_{b=1}^{N_i-1} C_{ab,i} = \frac{1}{N_i} \times \frac{1}{LIN_i} \times (N_i - 1) = \frac{1}{LIN_i}\left(1 - \frac{1}{N_i}\right), \qquad (2)$$

Therefore, the more institutions involved in paper $i$, the more RL mass contributed by the leading institution $a$. And the total RL flow intensity $C_{ab}$ from institution a to institution $b$ is calculated as

$$C_{ab} = \sum_{i=1}^{M_b} C_{ab,i}, \qquad (3)$$

where $M_b$ is the number of papers where $a$ is the leading institution and $b$ is a participating institution. And institution $a$'s total RL mass, total RL flow intensity to all other institutions is calculated as

$$LM_a = \sum_{b=1}^{B} C_{ab}, \qquad (4)$$

where $B$ is the number of institutions that institution $a$ has led.

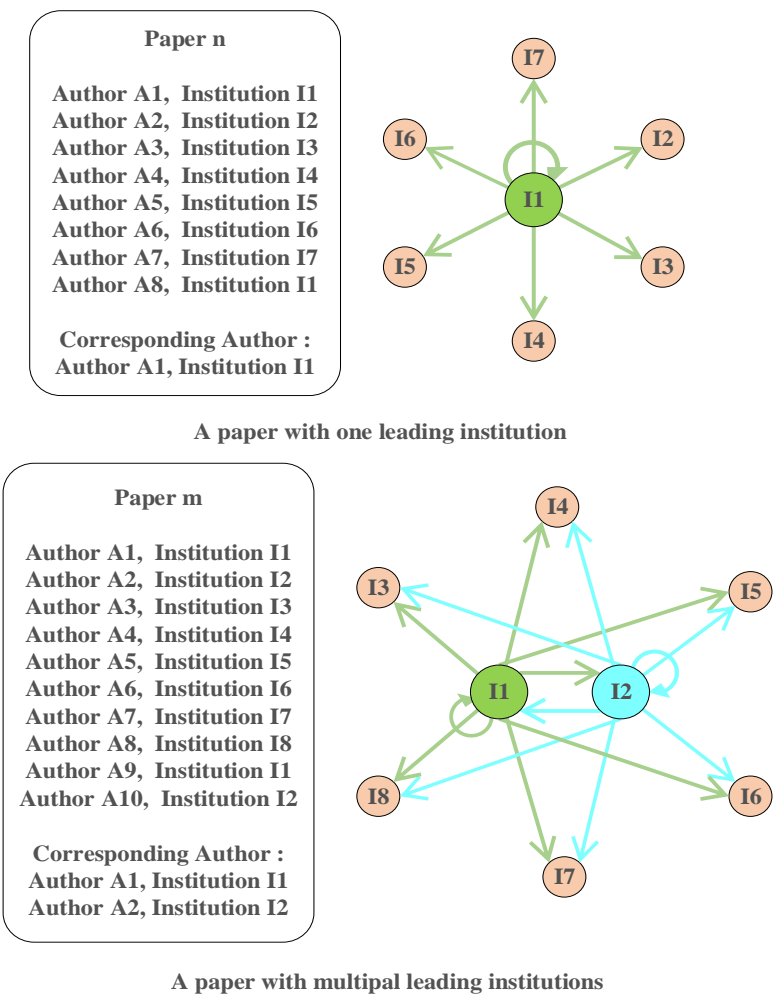

**Figure 1 The RL flow from the leading institution(s) to participating institutions**

## 3.2 Data

We performed a data collection in Web of Science Core Citation Database. More specifically, the data was retrieved using the search term CU = A AND SU = B AND PY = C, where A is "PEOPLES R CHINA", B is research areas in a certain field[2], and C is 2013-2017. Eventually, we obtained the complete authorship and collaboration information of 886105 articles published by all Chinese institutions during 2013-2017. Eventually, for "Life Sciences & Biomedicine" field, we sampled 484,903 articles published by 244 Chinese institutions, which have been the primary affiliation of the corresponding author for at least one paper (with multiple institutions) in each year. Similarly, we sampled 179,840 articles by 209 Chinese institutions for "Technology" field, 204,320 articles in "Physical Sciences" by 221 Chinese institutions, 13,628 articles in "Social Sciences" by 95 Chinese institutions, and 3,414 articles in "Humanities & Art" by 41Chinese institutes.

---

[2] http://images.webofknowledge.com/WOKRS532FR6/help/WOS/hp_research_areas_easca.html

### 3.3 Model and variables

To analyze the determinants of research leadership among different institutions, we adopt a gravity model. In its elementary form, the gravity model can be expressed as

$$C_{ij} = K \frac{M_i^{\beta_1} \times M_j^{\beta_2}}{d_{ij}^{\beta_3}}, \quad (5)$$

where $C_{ij}$ is the collaboration intensity (i.e., the number of co-publication) between institutions $i$ and $j$; $K$ is a proportionality constant; $M_i$ and $M_j$ are the numbers of publications of institutions $i$ and $j$; $d_{ij}$ is the geographical distance between the two institutions; $\beta_1$ and $\beta_2$ reflect the potential to collaboration and $\beta_3$ is an impedance factor reflecting the distance decaying factor for research collaboration. Taking logarithms of both sides and adding a random disturbance term, the multiplicative form of gravity model (1) can be converted into a testable linear stochastic model:

$$\ln C_{ij} = \ln K + \beta_1 \ln M_i + \beta_2 \ln M_j - \beta_3 \ln d_{ij} + \epsilon_{ij}, \quad (6)$$

And there has been a substantial body of literature involving the subject of which regression model are suitable to estimate (6) (Burger, Van Oort et al. 2009). Because RL of an institution reflects the extent to which it leads collaborations, and the aforementioned proximity measures quantify the distance between institutions, the gravity model is an appropriate modeling framework for our study.

Given the fractional count nature of the data and the existence of a large number of zeros (many institution pairs have no research collaboration), the ordinary least square (OLS) estimates with a censored dependent variable may be biased and inconsistent. Tobit regression, one of the limited dependent variable models (Li 2002), can effectively estimate linear relationship among variables when there was either left- or right-censoring in the dependent variables (Wooldridge 2003, Zhang and Lin 2018). In line with previous studies (Plotnikova and Rake 2014, Fernandez, Ferrandiz et al. 2016), we adopt a Tobit regression model where we consider zero leadership as left-censoring of the distribution.

To explore the roles and their dynamic evolutions of multiple proximities in shaping RL flows, we first conduct a cross-section estimate by the pooling data of 2013-2017, and then we perform cross-section estimates using two sub-period data. The equation that will be estimated is:

$$C_{ij} = \beta_0 + \beta_1 \ln(LM_i) + \beta_2 \ln(LM_j) + \beta_3 \ln(Geoprox_{ij}) + \beta_4 \ln(Cognprox_{ij}) + \beta_5 Instprox_{ij} + \beta_6 Socprox_{ij} + \beta_7 \ln(Econprox_{ij}) + \epsilon_{ij}, \quad (7)$$

The dependent variable $C_{ij}$ in our model is the RL flow intensity from institution $i$ (the leading institution) to institution $j$ during the period 2013-2017. In independent variables, time lags are used to avoid endogeneity and reverse causality (Fernandez, Ferrandiz et al. 2016, Zhang 2016, Gui, Liu et al. 2018). Specifically, the independent variables are lagged, capturing the information for 2008-2012. They are defined as follows,

The $LM_i$ and $LM_j$ refer to the RL mass of leading institution $i$ and participating institution $j$ respectively during 2008-2012.

Geographical proximity ($Geoprox_{ij}$) is the spatial distance between institution $i$ and institution $j$. It's calculated following the big circle formula with the latitude and longitude of the institution from Google Map (Gui, Liu et al. 2018).

Cognitive proximity represents the extent of overlap or the closeness of researchers' knowledge. In this paper, we adopt the Latent Dirichlet Allocation (LDA) model, to categorize the papers into 50 topics based on the keywords of each paper. And then we embed all the institutions into a 50-dimensional feature vector according to their publications during 2008-2012. The cognitive proximity between two institutions is calculated as the cosine similarity of two corresponding feature vectors[3].

Institutional proximity ($Instprox_{ij}$) is defined by the similarity on both formal and informal laws and practice. China's research resource and policy is distributed at the provincial level. In this paper, institutional proximity is a dummy variable (Lander 2015), which equals to 1 if institutions $i$ and $j$ are in the same province, and 0 otherwise. Note that in our data, there is little collinearity concern between $Instprox_{ij}$ and $Geoprox_{ij}$.

---

[3] Supplementary material provides a details of measurement of cognitive proximity.

Social proximity ($Socprox_{ij}$) draws on the social embeddedness based on the more intangible closeness, e.g., friendship or prior collaboration experience between researchers. Here, social proximity is a dummy variable, which equals to 1 if $C_{ij}$ >0 or $C_{ji}$ >0 during 2008-2012, and 0 otherwise (Fernandez, Ferrandiz et al. 2016).

Economic proximity ($Econprox_{ij}$) indicates the difference in academic resources between institutions $i$ and $j$. Acosta and Coronado used the difference in the total amount of R&D to measure the economic proximity between two entities (Acosta, Coronado et al. 2011). In this paper, we measure the $Econprox_{ij}$ as the absolute difference in the number of National Natural Science Foundation of China (NSFC) projects secured by institutions $i$ and $j$ during 2008-2012.[4]

---

[4] http://www.nsfc.gov.cn/

**Table 2. Description of dependent variables and independent variables**

| Variable | Description | Source |
|---|---|---|
| $C_{ij}$ | RL flow intensity from $i$ (leading institution) to $j$ (participating institution) in the period 2013-2017 | Web of Science |
| $LM_i$ | RL mass of institution $i$ (leading institution) in period 2008-2012 (Variable in logarithms) | Web of Science |
| $LM_j$ | RL mass of institution $j$ (participating institution) in period 2008-2012 (Variable in logarithms) | Web of Science |
| $Geoprox_{ij}$ | Geographical distance between institution $i$ and $j$, in kilometers (Variable in logarithms) | Google Map |
| $Cognprox_{ij}$ | Cosine similarity between institution vector pairs in period 2008-2012 (Variable in logarithms) | Web of Science |
| $Instprox_{ij}$ | Dummy variable, which take value 1 when institution $i$ and $j$ are in the same province | Google Map |
| $Socprox_{ij}$ | Dummy variable, which takes value 1 if institution $i$ and $j$ have collaborated in period 2008-2012 | Web of Science |
| $Econprox_{ij}$ | Absolute difference in the number of Natural Science Foundation Project of institution $i$ and $j$ in period 2008-2012 (Variable in logarithms) | NSF, China |

## 4 Results

For simplicity, we mainly introduce the results in the field of "Life sciences & biomedicine". We introduce the difference of the results in other fields in sub-section 4.4.

### 4.1 Descriptive analysis

For contextualization, we present the top 15 institutions by RL mass ($LM_i$) and top 15 institution pairs by RL flow intensity ($C_{ab}$) in Table 3 and Table 4, respectively. And Figure 2 and 3 show the spatial pattern and the dynamic pattern of RL flows, respectively.

As shown in Table 3, all these top ranked institutions (with large RL mass) are famous universities or institutions in biological and medical research in China. In particular, the top four institutions are the only four Chinese universities ranked top 200 in the field of Clinical Medicine and Pharmacy in Academic Ranking of World Universities 2016 edition[5]. These 15 institutions account for 33.41% of total RL among the 244 Chinese institutions. It's worth noting that the rank here only measures the RL in the collaborations with Chinese institutions. The RL in international collaborations is not accounted.

Table 3 Top 15 institutions by leadership mass

| Institution | Leadership Mass ($LM$) | $LM$/Total (%) | Acum (%) |
| --- | --- | --- | --- |
| Shanghai Jiao Tong Univ | 7386.08 | 3.99 | 3.99 |
| Peking Univ | 7077.45 | 3.82 | 7.81 |
| Sun Yat Sen Univ | 6344.97 | 3.43 | 3.82 |
| Fudan Univ | 6313.76 | 3.41 | 7.25 |
| Zhejiang Univ | 5904.04 | 3.19 | 10.66 |
| Shandong Univ | 5785.08 | 3.13 | 13.85 |
| Capital Med Univ | 4663.21 | 2.52 | 16.98 |
| Nanjing Med Univ | 4475.98 | 2.42 | 19.50 |
| Chinese Acad Med Sci | 4359.86 | 2.36 | 21.91 |
| Sichuan Univ | 3600.02 | 1.94 | 24.27 |
| Chinese Acad Agr Sci | 3450.40 | 1.86 | 26.21 |
| Huazhong Univ Sci & Technol | 3377.42 | 1.82 | 28.08 |
| Peoples Liberat Army Gen Hosp | 3331.95 | 1.80 | 29.90 |
| Cent S Univ | 3155.35 | 1.70 | 31.70 |
| Second Mil Med Univ | 3100.68 | 1.68 | 33.41 |
| Others | 123267.94 | 66.59 | 100 |
| Total | 185114.79 | 100 | |

---

5  http://www.shanghairanking.com/FieldMED2016.html

Table 4 shows the top-15 institution pairs by RL flow intensity from the leading institution to participating institution. A clear reciprocal pattern presents. Except for the RL flow from "China Agr Univ" to "Chinese Acad Agr Sci", each RL flow its reciprocal form among top 15. For example, the RL flow intensity from "Fudan Univ" to "Shanghai Jiao Tong Univ" is the largest, and the RL flow intensity from "Shanghai Jiao Tong Univ" to "Fudan Univ" is the second largest. Furthermore, all the top 15 institution pairs consist of institutions in the same city.

**Table 4 Top 15 Institution pairs by leadership intensity**

| Leading institution | Leading city | Participating institution | Participating city | RL flow intensity |
|---|---|---|---|---|
| Fudan Univ | Shanghai | Shanghai Jiao Tong Univ | Shanghai | 646.47 |
| Shanghai Jiao Tong Univ | Shanghai | Fudan Univ | Shanghai | 601.55 |
| Sun Yat Sen Univ | Guangzhou | Guangzhou Med Univ | Guangzhou | 377.53 |
| Shanghai Jiao Tong Univ | Shanghai | Tongji Univ | Shanghai | 351.73 |
| Peking Univ | Beijing | Capital Med Univ | Beijing | 324.11 |
| Nanjing Univ | Nanjing | Nanjing Med Univ | Nanjing | 315.64 |
| Tongji Univ | Shanghai | Shanghai Jiao Tong Univ | Shanghai | 297.83 |
| Second Mil Med Univ | Shanghai | Shanghai Jiao Tong Univ | Shanghai | 274.93 |
| Fudan Univ | Shanghai | Tongji Univ | Shanghai | 270.54 |
| Shanghai Jiao Tong Univ | Shanghai | Second Mil Med Univ | Shanghai | 261.11 |
| Nanjing Med Univ | Nanjing | Nanjing Univ | Nanjing | 259.17 |
| Capital Med Univ | Beijing | Peking Univ | Beijing | 257.21 |
| China Agr Univ | Beijing | Chinese Acad Agr Sci | Beijing | 237.37 |
| Tongji Univ | Shanghai | Fudan Univ | Shanghai | 224.59 |
| Guangzhou Med Univ | Guangzhou | Sun Yat Sen Univ | Guangzhou | 209.28 |

Figure 2 sheds light on the spatial pattern of RL flows on province level. Generally, the RL flows are mainly concentrated on the eastern part of China, consistent with China's economic and population distributions. Specifically, the most prominent RL flows are from/to Beijing and Shanghai, indicating that many institutions have a RL relationship with institutions in major cities. Although the weights on many edges are small, indicating the low frequency of collaborations between the two provinces, most provinces have RL relationships with each other.

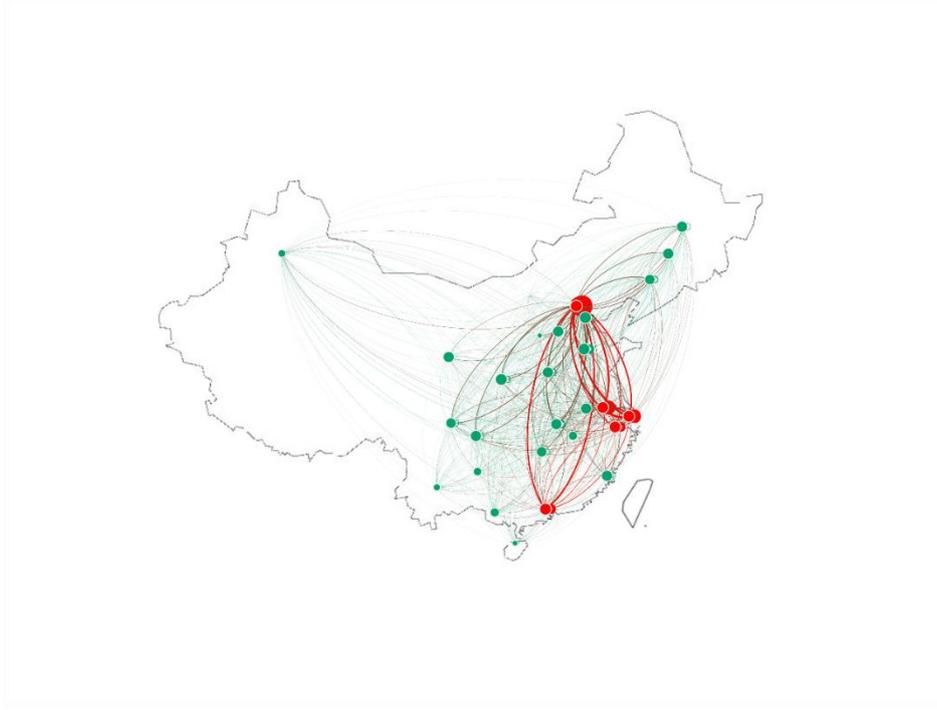

**Figure 2 Spatial pattern of RL flows at the province level**

Because of the existence of the proximities, the intensity of RL could be heterogeneously distributed. For example, institutions in the same city usually have higher chance to collaborate (Table 4). We adopt the disparity metric to measure the heterogeneity of RL intensity distribution. The disparity of institution $i$ is calculated as follows,

$$Disparity_i = \frac{(N_i - 1)\sum_{j=1,j\neq i}^{N_i}\left(\frac{C_{ij}}{LM_i}\right)^2 - 1}{N_i - 2}, \qquad (8)$$

where $N_i$ is the number of institutions that institution $i$ has led. If institution $i$'s RL is evenly distributed to all its participating institutions, $Disparity_i \to 0$. The more unevenly the institution $i$'s RL is distributed, the larger the $Disparity_i$ is. In the extreme case, when almost all of $i$'s RL flow to one participating institution, $Disparity_i \to 1$.

Figure 3 shows the kernel density distribution of institutions' leadership disparity during 2013-2017. In general, most institutions' disparity is small, ranging from 0 to 0.2. There exist a few institutions with very high disparity. Over these five years, the kernel density distribution changed steadily. The disparity of many institutions decreased, resulting in higher density around 0.1 and lower density over 0.2. This indicates that the

RL flow intensity has become more evenly distributed. A possible explanation is that the hindering effect of proximities has declined over time. We will discuss this in Section 4.3 Estimation for different sub-periods.

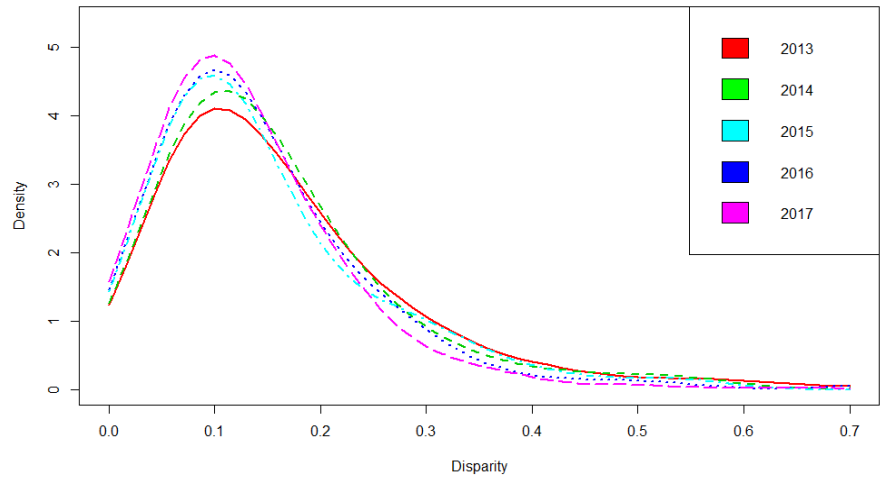

**Figure 3. Kernel density distribution of institutions' RL disparity over time**

Table 5 presents the descriptive statistics and the correlation matrix between variables. All variables variance inflation factors (VIFs) are lower than 3, indicating that there is no significant multicollinearity in the data.

**Table 5 Descriptive statistics and correlations**

| Variables | Mean | SD | VIF | 1 | 2 | 3 | 4 | 5 | 6 | 7 | 8 |
|---|---|---|---|---|---|---|---|---|---|---|---|
| $C_{ij}$ | 1.30 | 9.69 | - | 1 | | | | | | | |
| $ln(LM_i)$ | 4.84 | 1.27 | 1.61 | 0.18*** | 1 | | | | | | |
| $ln(LM_j)$ | 4.84 | 1.27 | 1.57 | 0.15*** | -0.00 | 1 | | | | | |
| $ln(Geoprox_{ij})$ | 6.59 | 1.24 | 2.33 | -0.17*** | -0.04*** | -0.04*** | 1 | | | | |
| $ln(Cognprox_{ij})$ | -0.07 | 0.04 | 1.63 | 0.15*** | 0.43*** | 0.43*** | 0.03*** | 1 | | | |
| $Instprox_{ij}$ | 0.07 | 0.25 | 2.34 | 0.18*** | 0.02*** | 0.02*** | -0.75*** | -0.04*** | 1 | | |
| $Socprox_{ij}$ | 0.13 | 0.34 | 1.31 | 0.28*** | 0.35*** | 0.28*** | -0.18*** | 0.31*** | 0.18*** | 1 | |
| $ln(Econprox_{ij})$ | 4.54 | 1.70 | 1.37 | 0.10*** | 0.36*** | 0.36*** | -0.03*** | 0.32*** | -0.01*** | 0.23*** | 1 |

*p<0.10; **p<0.05; ***p<0.01

### 4.2 Estimation result

In order to measure the determinants of research leadership, we also used Tobit gravity model to estimate the impacts. Table 6 reports the estimation results of our Tobit regression-based gravity model. Model 1 is the base gravity model, which only includes $LM_i$, $LM_j$, and $Geoprox_{ij}$. Models 2 to 5 have incrementally added independent variables. Model 5 presents the full model with all variables.

The positive and significant coefficient of $LM_i$ and $LM_j$ indicates that the RL mass of both the leading and participating institutions are positively associated with the RL flow intensity. Previous studies showed that the counts of publications of two institutions are positive determinants of their co-publication (Hoekman, Frenken et al. 2010, Parreira, Machado et al. 2017). Our results extended the understanding of the collaboration patterns between institutions by showing that the institutions with rich leading experience are more likely to lead future research, as well as to participate in others' research. In addition, we found that the coefficients of $ln(LM_i)$ are larger than that of $ln(LM_j)$ in all models, indicating that there is a larger influence of the leading institution's RL mass on RL flow intensity.

**Table 6 Estimation results of Tobit gravity model**

|  | Model (1) | Model (2) | Model (3) | Model (4) | Model (5) |
|---|---|---|---|---|---|
| $ln(LM_i)$ | 8.440*** | 5.963*** | 5.999*** | 4.551*** | 4.417*** |
|  | (0.105) | (0.114) | (0.114) | (0.116) | (0.124) |
| $ln(LM_j)$ | 7.024*** | 4.433*** | 4.453*** | 3.360*** | 3.232*** |
|  | (0.101) | (0.114) | (0.113) | (0.114) | (0.122) |
| $ln(Geoprox_{ij})$ | -4.246*** | -4.606*** | -1.963*** | -1.600*** | -1.590*** |
|  | (0.087) | (0.088) | (0.143) | (0.142) | (0.142) |
| $ln(Cognprox_{ij})$ |  | 237.725*** | 241.254*** | 200.344*** | 201.403*** |
|  |  | (5.894) | (5.889) | (5.76) | (5.783) |
| $Instprox_{ij}$ |  |  | 15.371*** | 12.547*** | 12.648*** |
|  |  |  | (0.679) | (0.671) | (0.673) |
| $Socprox_{ij}$ |  |  |  | 13.224*** | 13.210*** |
|  |  |  |  | (0.326) | (0.327) |
| $ln(Econprox_{ij})$ |  |  |  |  | 0.277*** |
|  |  |  |  |  | (0.089) |
| _cons | -71.408*** | -28.875*** | -47.347*** | -41.155*** | -41.178*** |
|  | (1.027) | (1.352) | (1.591) | (1.568) | (1.569) |
| N |  |  | 59292 |  |  |
| Left-censored |  |  | 46,757 |  |  |
| Uncensored |  |  | 12,535 |  |  |
| LR chi2 | 14871.14 | 16827.87 | 17331.69 | 19000.90 | 19010.52 |
| Log-likelihood | -66308.096 | -65329.73 | -65077.822 | -64243.217 | -64238.407 |
| Pseudo R2 | 0.1008 | 0.1141 | 0.1175 | 0.1288 | 0.1289 |
| Prob > chi2 | 0.0000 | 0.0000 | 0.0000 | 0.0000 | 0.0000 |

*$p<0.10$; **$p<0.05$; ***$p<0.01$

The negative and significant coefficient of $Geoprox_{ij}$ suggests that RL flow intensity decays with longer geographical distance. This is in line with previous literature on research collaboration (Berge 2017, Parreira, Machado et al. 2017). Coordination activities such as seminars, meetings, exchange of personnel, and sharing lab facilities become more difficult and expensive as distance increases. In addition, successful collaborative research projects involve intensive face-to-face discussions, which are difficult if the two institutions are far away from each other.

Cognitive proximity is positively significant, indicating that RL flows are more likely to occur between institutions with similar research experience. This echoes the previous findings that researchers need a shared cognitive base to understand, absorb, and explore the unknown successfully Boschma (2005).

The institutional proximity is positively significant, indicating that positive and significant, showing that factors such as similar policies and culture background could facilitate the RL flows. This result is in line with existing studies (Fernandez, Ferrandiz et al. 2016), and can also explain the phenomenon that the leading institution pairs are mostly located in the same city.

Prior collaborations are found to be positively associated with the RL flows. This result indicates that prior collaborations lead to higher trust and confidence in future collaborations, and is in line with the previous studies (Lazzeretti and Capone 2016).

The positive and significant coefficient of economic proximity suggests that RL flows are more likely to occur between institutions with a diverse academic economic resources. This is consistent with the center-periphery hypothesis (Schott 1998) that researchers in peripheral regions are willing to collaborate with those in core regions to gain access to research resources, while core region researchers are also willing to seek complementarities to participate in their research (Hwang 2008).

**4.3 Estimation for different sub-periods**

To analyze the determinants of RL flows from a dynamic perspective, we estimate the full model for two sub-periods (2013-2014 and 2016-2017) and compare their parameters over time. We take 2-year lagged independent variables to address the endogeneity and reverse causality concerns (Fernandez, Ferrandiz et al. 2016, Zhang 2016, Gui, Liu et al. 2018). Table 7 presents the estimation results for different sub-periods. To compare the fitted model in different sub-years, we adopt the Chow test (Chow 1960) to determine whether the independent variables have significant differences in the sub-periods. The Chow test result rejects no difference specification (p<0.01), indicating a clear difference between the two models for two sub-periods.

**Table 7 Estimation results of Tobit gravity model for sub-period**

|  | Model (1) | Model (2) |
|---|---|---|
|  | 2013-2014 | 2016-2017 |
| $ln(LM_i)$ | 2.181*** | 2.121*** |
|  | (0.079) | (0.062) |
| $ln(LM_j)$ | 1.615*** | 1.784*** |
|  | (0.077) | (0.060) |
| $ln(Geoprox_{ij})$ | -0.857*** | -0.797*** |
|  | (0.094) | (0.078) |
| $ln(Cognprox_{ij})$ | 83.603*** | 84.956*** |
|  | (4.382) | (3.649) |
| $Instprox_{ij}$ | 4.592*** | 4.970*** |
|  | (0.444) | (0.366) |
| $Socprox_{ij}$ | 14.668*** | 12.034*** |
|  | (0.225) | (0.181) |
| $ln(Econprox_{ij})$ | 0.069 | 0.102*** |
|  | (0.063 | (0.051) |
| _cons | -27.354*** | -27.994*** |
|  | (0.953) | (0.775) |
| N | 59,292 | 59,292 |
| Left-censored | 52,142 | 50,518 |
| Uncensored | 7,150 | 8,774 |
| LR chi2 | 19219.97 | 22033.39 |
| Prob > chi2 | 0.0000 | 0.0000 |
| Pseudo R2 | 0.2320 | 0.2263 |
| Log likelihood | -31817.755 | -37667.064 |
| Chow test | 43.91*** | |

*$p<0.10$; ** $p<0.05$; *** $p<0.01$*

Consistent with the model fitted with all data (Table 6), the coefficients for $ln(LM_i)$

and $ln(LM_j)$ are both positively significant in both sub-periods. More specifically, although the coefficient of the $ln(LM_i)$ have been larger than that of $ln(LM_j)$, the coefficient of $ln(LM_i)$ decreased, while the coefficient of $ln(LM_j)$ increased over time. This indicates that the RL of the leading institution has a more prominent effect than that of the participating institution on the future RL flow, but the difference has been becoming smaller over time.

The geographical proximity remains negatively significant, while its absolute value of the negative coefficient decreased, indicating that the negative effect of geographical proximity has declined over time. On the other hand, the cognitive proximity, and institutional proximity remain positively significant and their coefficients increased, indicating that RL flows have become more likely to occur between institutions that have a similar research background and are located in the same province. The social proximity is positively significant in both sub-periods, but its coefficient declines, indicating that although the prior collaborations enhance the chance of future RL flows, its influence is decreasing over time.

It's interesting that the economic proximity was statistically insignificant in 2013-2014, and became positively significant in 2016-2017. This result indicates that the economic proximity has recently become an important determinant of RL flows; institutions have become increasingly likely to collaborate if there is a gap in economic resource (measured by the number of NSFC projects). The center–periphery theory provides a possible explanation for this: the institutions with fewer economic resource tend to participate in the projects led by those with more resources; institutions with much economic resource are willing to collaborate with those less resource-intensive institutions for the access to non-economic research resources (e.g. patient subjects, the habitat for certain animals/plants, human resources, etc.).

**4.4 The role of proximity on RL in multiple research fields**

To have a comprehensive understanding of RL, we also exam the role of proximity on

RL in different research fields. We combined "Humanities & Arts" and "Social Sciences" into "Humanities, Arts and Social Sciences" (HASS) because they have been found to have similar collaboration pattern, outcome form and collaborative scale (Engels, Ossenblok et al. 2012). Existing literature usually combined these two fields to explore the collaboration patterns (Sidone, Haddad et al. 2017). In addition, the number of publications in these two fields are far fewer than that of natural sciences and medicine by Chinese institutions. Combining them can avoid unstable results caused by the unbalanced data. Table 8 shows the estimation results of Tobit gravity model in these fields during 2013-2017. Generally, the results are in line with that of "Life Sciences & Biomedicine".

Table 9 presents the estimation results of Tobit gravity model for sub-period in other fields. For these fields, the RL mass of both leading and participating institutions remain a positive and significant coefficient. And the gap is narrowing over time. Different from the observations in "Life Sciences & Biomedicine", both effect of RL mass of leading and participating institutions are decreasing. The result of geographical proximity, cognitive proximity and social proximity are in line with that of "Life Sciences & Biomedicine". As for the institutional proximity and economic proximity, the results of Physical Sciences and Technology remain the same as that of "Life Sciences & Biomedicine". However, in HASS, these two coefficients are declining, indicating the hinder effect of disparity in policy, norms, rules, and economics has been becoming less significant in HASS RL flows.

**Table 8 Estimation results of Tobit gravity model in other fields**

|  | Technology | Physical Sciences | HASS |
|---|---|---|---|
|  | 2013-2017 | 2013-2017 | 2013-2017 |
| $ln(LM_i)$ | 1.493*** | 1.356*** | 0.497*** |
|  | (0.026) | (0.023) | (0.046) |
| $ln(LM_j)$ | 1.348*** | 1.267*** | 0.411*** |
|  | (0.025) | (0.023) | (0.046) |
| $ln(Geoprox_{ij})$ | -0.454*** | -0.446*** | -0.289*** |
|  | (0.026) | (0.026) | (0.055) |
| $ln(Cognprox_{ij})$ | 18.452*** | 17.177*** | 5.805*** |
|  | (0.982) | (0.982) | (0.590) |
| $Instprox_{ij}$ | 2.210*** | 2.237*** | 1.981*** |
|  | (0.047) | (0.044) | (0.011) |
| $Socprox_{ij}$ | 2.043*** | 2.157*** | 0.762** |
|  | (0.116) | (0.108) | (0.244) |
| $ln(Econprox_{ij})$ | 0.031* | 0.099*** | 0.172*** |
|  | (0.017) | (0.016) | (0.034) |
| _cons | -3.851*** | -4.093*** | -0.488*** |
|  | (0.215) | (0.204) | (0.420) |
| N | 43,472 | 48,620 | 8,930 |
| Left-censored | 30,646 | 34,040 | 7,138 |
| Uncensored | 12,826 | 14,580 | 1,792 |
| LR chi2 | 24880.19 | 26879.96 | 2 2451.48 |
| Log-likelihood | -40594.615 | -46303.218 | -6093.6622 |
| Pseudo R2 | 0.2346 | 0.2250 | 0.1675 |
| Prob > chi2 | 0.0000 | 0.0000 | 0.0000 |

*$p<0.10$; **$p<0.05$; ***$p<0.01$

**Table 9 Estimation results of Tobit gravity model for sub-period in other fields**

|  | Technology | | Physical Sciences | | HASS | |
|---|---|---|---|---|---|---|
|  | 2013-2014 | 2016-2017 | 2013-2014 | 2016-2017 | 2013-2014 | 2016-2017 |
| $ln(LM_i)$ | 0.916*** | 0.668*** | 0.864*** | 0.642*** | 0.570*** | 0.431*** |
|  | (0.018) | (0.015) | (0.016) | (0.015) | (0.037) | (0.049) |
| $ln(LM_j)$ | 0.792*** | 0.630*** | 0.762*** | 0.616*** | 0.474*** | 0.409*** |
|  | (0.018) | (0.015) | (0.016) | (0.015) | (0.036) | (0.049) |
| $ln(Geoprox_{ij})$ | -0.249*** | -0.223*** | -0.224*** | -0.203*** | -0.178*** | -0.091*** |
|  | (0.020) | (0.018) | (0.018) | (0.017) | (0.044) | (0.058) |
| $ln(Cognprox_{ij})$ | 15.689*** | 19.765*** | 13.555*** | 15.827*** | 4.946*** | 6.404*** |
|  | (0.987) | (0.705) | (0.793) | (0.705) | (0.666) | (0.608) |
| $Instprox_{ij}$ | 1.637*** | 1.656*** | 1.634*** | 1.696*** | 1.555*** | 1.335*** |
|  | (0.036) | (0.031) | (0.033) | (0.030) | (0.109) | (0.074) |
| $Socprox_{ij}$ | 1.171*** | 1.149*** | 1.123*** | 1.008*** | 0.947*** | 0.373* |
|  | (0.085) | (0.079) | (0.077) | (0.075) | (0.257) | (0.194) |
| $ln(Econprox_{ij})$ | 0.002* | 0.049*** | 0.031** | 0.044*** | 0.079** | 0.051** |
|  | (0.012) | (0.011) | (0.011) | (0.011) | (0.035) | (0.026) |
| _cons | -3.084*** | -4.923*** | -3.387*** | -4.706*** | -2.483*** | -2.092*** |
|  | (0.165) | (0.152) | (0.149) | (0.150) | (0.455) | (0.328) |
| N | 43,472 | 43,472 | 48,620 | 48,620 | 8,930 | 8,930 |
| Left-censored | 36,936 | 33,150 | 40,807 | 37,359 | 8250 | 7610 |
| Uncensored | 6,536 | 10,322 | 7,813 | 11,261 | 680 | 1,320 |
| LR chi2 | 15342.68 | 22533.58 | 17854.47 | 23926.17 | 1300.24 | 2175.00 |
| Prob > chi2 | 0.0000 | 0.0000 | 0.0000 | 0.0000 | 0.0000 | 0.0000 |
| Pseudo R2 | 0.2764 | 0.2786 | 0.2747 | 0.2712 | 0.2126 | 0.2035 |
| Log likelihood | -20088.156 | -29176.044 | -23568.254 | -32147.945 | -2408.0288 | -4256.5959 |
| Chow test | 285.18*** | | 254.35*** | | 26.69*** | |

*$p<0.10$; **$p<0.05$; ***$p<0.01$

Figure 4.a, 4.b, 4.c show the kernel density distributions of institutions' RL disparity during 2013-2017 in "Technology", "Physical Sciences", and HASS. The results are in line with Figure 3, indicating that the RL flow intensity has become more evenly distributed. It is worth noting that the change of HASS is more pronounced, which is consistent with the discussions above.

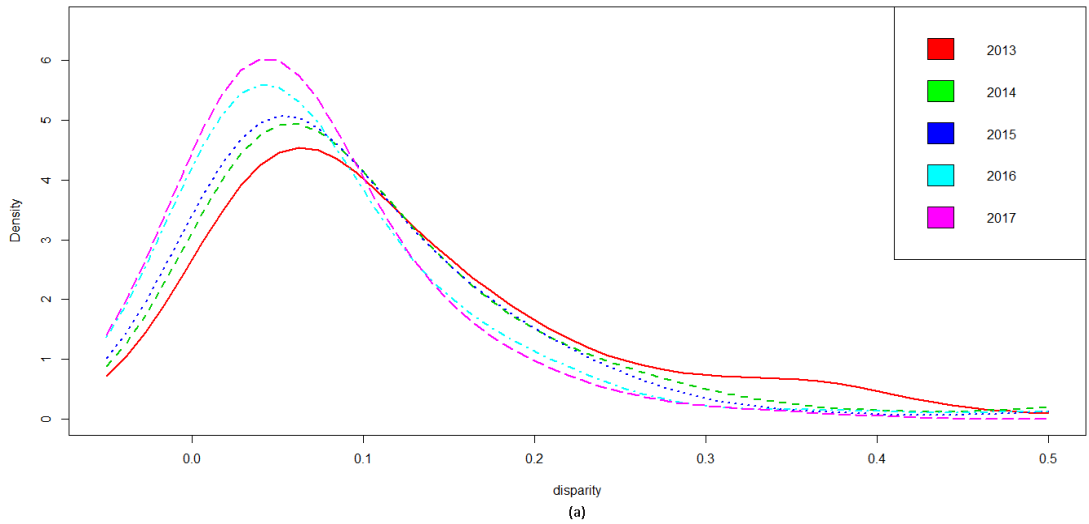

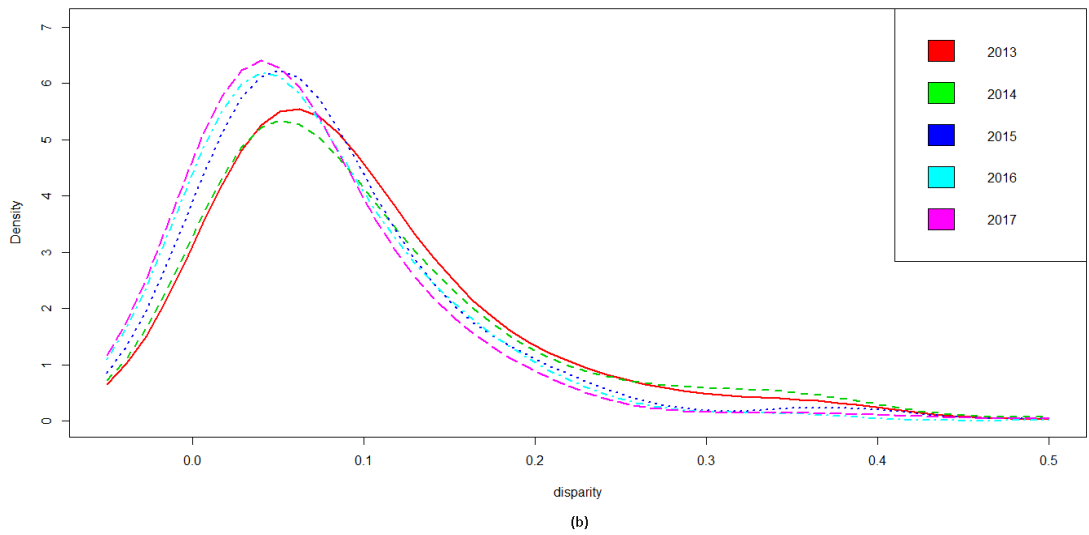

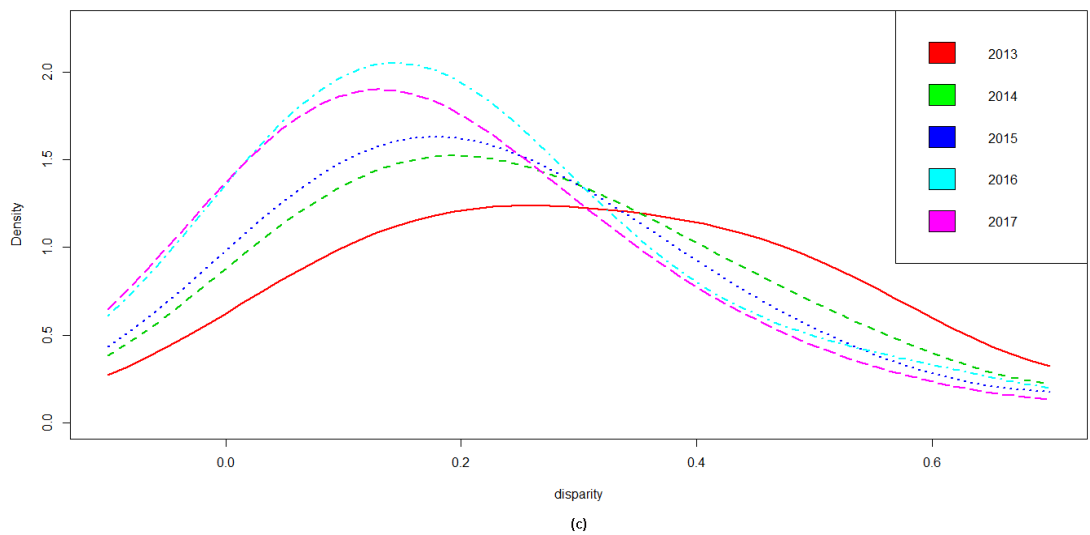

**Figure 4 Kernel density distributions of institutions' RL disparity over time Technology (a), Physical Sciences (b), HASS (c)**

# 5 Robustness check

To further test the robustness of our results, we apply the "full counting" method (i.e. the count of participating institutions) to measure the RL flow intensity and then estimate a negative binomial regression for the data of "Life sciences & biomedicine". For example, in a paper with three institutions, the leading institution's RL flow to each of the two participating institutions is 1. The RL mass $LM_i$ obtained by the leading institution is, therefore 2. Since the dependent variable is count data with over-dispersion (i.e. its variance is greater than its mean) and there exist a large number of zeros, a zero-inflated negative binomial regression is adopted (Cameron and Trivedi 2013). The rest of the independent variables remain the same as the main model. The results of the zero-inflated negative binomial regression (Table 10) leads to the same conclusions and confirms the robustness of the main model in Table 6.

**Table 10 Estimation results of zero-inflated negative binomial regression ("Full counting" of RL flows and mass)**

|  | Model (1) | Model (2) | Model (3) | Model (4) | Model (5) |
|---|---|---|---|---|---|
| $ln(LM_i)$ | 0.578*** | 0.428*** | 0.439*** | 0.287*** | 0.307*** |
|  | (0.010) | (0.011) | (0.011) | (0.011) | (0.012) |
| $ln(LM_j)$ | 0.455*** | 0.322*** | 0.330*** | 0.232*** | 0.249*** |
|  | (0.010) | (0.010) | (0.010) | (0.010) | (0.011) |
| $ln(Geoprox_{ij})$ | -0.343*** | -0.353*** | -0.137*** | -0.091*** | -0.092*** |
|  | (0.008) | (0.007) | (0.012) | (0.012) | (0.012) |
| $ln(Cognprox_{ij})$ |  | 13.896*** | 14.804*** | 12.722*** | 12.208*** |
|  |  | (0.564) | (0.565) | (0.547) | (0.555) |
| $Instprox_{ij}$ |  |  | 1.230*** | 0.980*** | 0.983*** |
|  |  |  | (0.058) | (0.057) | (0.0566) |
| $Socprox_{ij}$ |  |  |  | 1.271*** | 1.260*** |
|  |  |  |  | (0.029) | (0.028) |
| $ln(Econprox_{ij})$ |  |  |  |  | -0.030*** |
|  |  |  |  |  | (0.008) |
| _cons | -2.925*** | -0.544*** | -2.150*** | -1.660*** | -1.728*** |
|  | (0.097) | (0.127) | (0.149) | (0.145) | (0.145) |
| *Inflated (Logit)* |  |  |  |  |  |
| $ln(LM_i)$ | -0.782*** | -0.517*** | -0.520*** | -0.433*** | -0.371*** |
|  | (0.018) | (0.018) | (0.018) | (0.021) | (0.023) |
| $ln(LM_j)$ | -0.663*** | -0.381*** | -0.383*** | -0.310*** | -0.252*** |
|  | (0.017) | (0.017) | (0.018) | (0.020) | (0.021) |
| $ln(Geoprox_{ij})$ | 0.277*** | 0.317*** | 0.131*** | 0.084*** | 0.077*** |
|  | (0.015) | (0.015) | (0.023) | (0.026) | (0.026) |
| $ln(Cognprox_{ij})$ |  | -23.581*** | -23.645*** | -19.337*** | -20.168*** |
|  |  | (0.861) | (0.881) | (0.933) | (0.938) |
| $Instprox_{ij}$ |  |  | -1.286*** | -0.979*** | -0.996*** |
|  |  |  | (0.110) | (0.123) | (0.122) |
| $Socprox_{ij}$ |  |  |  | -1.985*** | -2.017*** |
|  |  |  |  | (0.110) | (0.111) |
| $ln(Econprox_{ij})$ |  |  |  |  | -0.104*** |
|  |  |  |  |  | (0.015) |
| _cons | 6.021*** | 1.542*** | 2.864*** | 2.725*** | 2.597*** |
|  | (0.156) | (0.208) | (0.253) | (0.281) | (0.280) |
| Lnalpha | 0.837*** | 0.640*** | 0.597*** | 0.387*** | 0.384*** |
| Alpha | 2.311 | 1.897 | 1.817 | 1.473 | 1.468 |
| Likelihood ratio test | 1.1e+05 | 8.9e+04 | 8.6e+04 | 7.9+04 | 7.9e+04 |
| Vuong test | 19.26 | 13.84 | 14.84 | 19.24 | 19.03 |
| LR chi2 | 7960.96 | 8744.57 | 9221.41 | 11697.79 | 11712.96 |
| Number of obs |  |  | 59,292 |  |  |
| Nonzero obs |  |  | 12,585 |  |  |

*$p<0.10$; ** $p<0.05$; *** $p<0.01$

## 6 Conclusions

In this paper, we proposed the concept and measurement of "research leadership" (RL) in research collaborations and examined the determinants of RL flow between Chinese institutions using a comprehensive bibliographic dataset during 2013-2017. We find that research leadership among Chinese institutions is highly concentrated. Top 15 institutions account for over one-third of the total research leadership in the field of biomedical field. The most prevalent research leadership flows occur between institutions in the same city. In addition, the flows of research leadership from the leading institution are observed to be (increasingly) evenly distributed to multiple participating institutions.

Our empirical results drawn from a Tobit regression-based gravity model show that the research leadership mass of both the leading and participating institutions and the geographical, cognitive, institutional, social and economic proximities are important factors of the flow of research leadership among Chinese institutions.

These results remain robust to several sensitive checks. In particular, the leading institution's research leadership mass has a higher influence than that of the participating institution, though the gap is narrowing. The constraining effect of geographical and social barriers have become less significant. But cognitive proximity is playing increasingly important roles in science and technology fields. Notably, we also obtain evidence that economic proximity has recently become an important factor of research leadership flows in science and technology fields.

Combining the results from both the descriptive statistics and the gravity model, there is clear evidence that although there are a number of significant proximities for research leadership flows, the effects of these proximities have been declining recently. The collaborations among institutions become more "flat" and convenient, possibly due to the advances in transportation, communication, and the general research capability of Chinese institutions.

This research leads to the following policy implications. Given the positive signifi-cance of both the leading and participating institutions research leadership mass, institutions should actively lead research projects to obtain a more significant role in both leading and participating important research in the future. Given the hindering effect of geographical proximity social proximity and institutional proximity, the funding bodies should encourage cross-provincial collaborations. In the meanwhile, policymakers in different provinces should facilitate the research leadership flow across multiple provinces by unifying their research policy and norms. In addition, policy makers and funding bodies should facilitate the establishment of new collaborations between institutions that have not collaborated before, and between the institutions with rich economic resource and those with less economic resource. By doing so, the hindering effects of social and economic proximities among institutions are reduced, so that we can (a) take advantage of the network effect brought by additional links in the collaboration network, and (b) integrate the economic and complementary non-economic research resources possessed by different institutions。

Results of this study also shed light on future applications of the proposed research leadership concept in analyzing other scientific collaboration datasets, such as patent, grant, conference organizations, and journal editorial. It is also important to delve into the research leadership roles of individual scholars and the whole country.